\documentclass[onecolumn,%
notitlepage,%
nofootinbib,%
superscriptaddress,%
amsfonts,amsmath]{revtex4}

\begin{document}

\title{Rotating black holes in massive gravity}

\author{Eugeny~Babichev} 
\affiliation{Laboratoire de Physique Th\'eorique d'Orsay,
B\^atiment 210, Universit\'e Paris-Sud 11,
F-91405 Orsay Cedex, France}

\author{Alessandro Fabbri}
\affiliation{Museo Storico della Fisica e Centro Studi e Ricerche Enrico Fermi, Piazza del Viminale 1, 00184 Roma, Italy; Dipartimento di Fisica dell'Universit\`a di Bologna,
Via Irnerio 46, 40126 Bologna, Italy; Departamento de F\'isica Te\'orica and IFIC, Universidad de Valencia-CSIC, C. Dr. Moliner 50, 46100 Burjassot, Spain}

\begin{abstract}
We present a solution for rotating black holes in massive gravity. 
We first give a solution of massive gravity with one dynamical metric. 
Both metrics of this solution are expressed in the advanced Eddington-Finkelstein-like coordinates:
the physical metric has the original Kerr line element, while the fiducial metric is flat, but written in a rotating Eddington-Finkelstein form.
For the bi-gravity theory we give an analogue of this solution: the two metrics have the original Kerr form, 
but, in general, different black hole masses.
The generalisation of the solution to include the electric charge is also given, it is an analogue of the Kerr-Newman solution in General Relativity. 
We also discuss further possible ways to generalise the solutions.
\end{abstract}

\date{\today}


\maketitle

\section{Introduction}

Massive gravity is a modification of General Relativity, in which the graviton acquires a nonzero mass. 
The study of this theory is interesting from various points of view, see the recent review~\cite{deRham:2014zqa}.
It has been shown that in spite of the fact that the generic non-linear massive gravity model possesses the higher-derivative (Ostrogradski) ghost~\cite{Boulware:1973my},
there is a subfamily of mass terms for which the Boulware-Deser ghost does not appear~\cite{deRham:2010ik,Kluson:2011rt}, the de-Rham-Gabadadze-Tolley  (dRGT) model.
In this model the helicity-0 degree of freedom, which is still present in this massive gravity model, is effectively screened locally, thanks to the 
Vainshtein mechanism~\cite{Vainshtein:1972sx,Babichev:2009us,Koyama:2011xz,Babichev:2013usa}, thus restoring General Relativity. 

It is more difficult to find exact solutions in massive gravity theory than in General Relativity, due to the more complicated structure of the equations of motion. 
This is true, in particular, for black hole solutions. 
In the nonlinear massive gravity theory (with Boulware-Deser ghost) black hole solutions were found in~\cite{Salam:1976as}, although at that time 
this theory was studied in a completely different context. 
In the dRGT theory with one fixed Minkowski metric, 
a class of non-bidiagonal Schwarzschild-de-Sitter solutions was found in~\cite{Koyama:2011xz}.
In bi-gravity theory (with two dynamical metrics), spherically symmetric black hole solutions were found in~\cite{Comelli:2011wq,Volkov:2012wp} (see also \cite{Berezhiani:2008nr}).
For a particular choice of the parameters of the potential term spherically symmetric (charged and uncharged) 
solutions in dRGT model were given in~\cite{Nieuwenhuizen:2011sq,Berezhiani:2011mt}.
Later,  a general class of charged black hole solutions, in both 
dRGT model and its bi-gravity extension, was found in~\cite{Babichev:2014fka}, which includes  as particular cases the previously found spherically symmetric solutions.
A numerical solution of a black hole with a hair in dRGT model has been found in~\cite{Brito:2013xaa}.
See recent reviews on black holes in massive gravity~\cite{VolkovTasinato}.

Due to the presence of extra degrees of freedom, 
the stability properties of black holes in massive gravity is different from those of black holes in General Relativity. 
E.g., the bidiagonal Schwarzschild solutions in massive gravity have been shown 
to be unstable  to radial perturbations~\cite{Babichev:2013una,Brito:2013wya}, with the rate of instability of order of the graviton mass (for phenomenologically interesting situations). 
On the contrary, the non-bidiagonal solutions are stable with respect to the same type of perturbations~\cite{Babichev:2014oua}. 
There is a specific choice of parameters for which the perturbations are identical to those of General Relativity~\cite{Kodama:2013rea}, i.e such black holes are stable. 
For these black holes the dangerous radial mode is absent, and this fact 
is connected to the specific choice of the parameters of the Lagrangian~\cite{Babichev:2014oua}.

So far only spherically symmetric solutions have been found in massive gravity. 
One obvious reason is, as we mentioned, the complexity of the equations of motion
(one exception is two equal metrics:
in this case any solution of General Relativity is also a solution of 
massive gravity, since the massive potential terms in the metric equations are trivially zero).

In this paper we present, for the first time, a class of rotating black hole solutions in massive (bi)gravity.
We find solutions both for the dRGT theory with one dynamical metric and for its extension with two dynamical metrics.
It proves to be helpful to use the Eddington-Finkelstein-like coordinates for both metrics 
(the bi-Eddington-Finkelstein ansatz we used before to study perturbations of the non-bidiagonal solutions~\cite{Babichev:2013una} and 
to find charged black hole solutions~\cite{Babichev:2014fka}). 
Once we put forward the bi-Eddington-Finkelstein coordinates, 
the calculations greatly simplify and we are able to prove that 
the Kerr metric in these coordinates for the two metrics (with, in general, different masses) is indeed a solution of massive gravity equations, 
under some conditions on the conformal factor of one of the metrics and on the bare cosmological constant(s). 

\section{Rotating Kerr black hole solutions}

We first consider the original dRGT version of massive gravity~\cite{deRham:2010ik} with one dynamical metric $g$, and with fixed flat metric $f$,
\begin{eqnarray}\label{action}
S &=&  M^2_P\int d^4 x \sqrt{-g} \left(\frac{R[g]}{2} + m^2 \mathcal{U} [g,f] - m^2 \Lambda_g  \right) 	
\end{eqnarray}
where $R[g]$ is the Ricci scalar for the $g$-metric, $\Lambda_g$ 
is the (dimensionless) bare cosmological constant, $m$ is the mass parameter (related to the graviton mass),  and  $\mathcal{U}[g,f]$ is the potential term.
The interaction potential ${\cal U}[g,f]$ is expressed in terms of the matrix $\mathcal{K}^\mu_{\phantom{\mu}\nu} = \delta^\mu_\nu - \gamma^\mu_{\phantom{\mu}\nu}$, 
where $\gamma^\mu_{\phantom{\mu}\nu} = \sqrt{g^{\mu\alpha}f_{\alpha\nu}} $ and it  contains three parts, 
$$
	\mathcal{U} \equiv  \mathcal{U}_2 + \alpha_3 \mathcal{U}_3 + \alpha_4 \mathcal{U}_4,
$$
where $\alpha_3$ and $\alpha_4$ are parameters of the theory, and,
\begin{equation}
\label{U}
\begin{aligned}
	\mathcal{U}_2 &= \frac{1}{2!}\left( [\mathcal{K}]^2 - [\mathcal{K}^2] \right), \\
	\mathcal{U}_3 &= \frac{1}{3!}\left( [\mathcal{K}]^3 -3 [\mathcal{K}] [\mathcal{K}^2] +2[\mathcal{K}^3]\right), \\
	\mathcal{U}_4 &= \det (\mathcal{K}).
\end{aligned}
\end{equation}
where  $[\mathcal{K}]\equiv \mathcal{K}^\rho_{\phantom{\rho}\rho}$ and 
	$[\mathcal{K}^n]\equiv  (\mathcal{K}^n)^\rho_\rho $.

The metric equations of motion, obtained by variation of the action with respect to the metric $g$ read,
\begin{equation}
	G^{\mu}_{\phantom{\mu}\nu}  =  m^2\left( T^{\mu}_{\phantom{\mu}\nu} -  \Lambda_g \delta^\mu_\nu \right), \label{Eg}
\end{equation}
where $G^{\mu}_{\phantom{\mu}\nu}$ is the Einstein tensor built from the metric $g$,
and $T^{\mu}_{\phantom{\mu}\nu}$ is the mass energy-momentum tensor,
\begin{equation}\label{T}
T^{\mu}_{\phantom{\mu}\nu} \equiv \mathcal{U} \delta^{\mu}_{\nu} - 2 g^{\mu\alpha}\frac{\delta \mathcal{U}}{\delta g^{\nu\alpha}}.
\end{equation}

We assume $g$ to be the Kerr metric, written in the original form found by Kerr~\cite{Kerr:1963ud},
\begin{equation}
\begin{aligned}\label{g}
	ds_g^2 = 
	& -\left( 1- \frac{r_g r}{\rho^2}\right)\left( dv+a\sin^2\theta d\phi \right)^2 \\
			& + 2\left( dv+a\sin^2\theta d\phi \right)\left( dr+a\sin^2\theta d\phi \right) 
			+ \rho^2 \left( d\theta^2 +\sin^2\theta d\phi^2 \right).
\end{aligned}
\end{equation}
where
\begin{equation}
	\rho^2 = r^2+ a^2\cos^2\theta,
\end{equation}
and $r_g$ is the Schwarzschild radius of the black hole and $a$ is the rotation parameter. 

Note that these coordinates can be thought of  as an extension of the Eddington-Finkelstein coordinates for rotating space-time. 
The Eddington-Finkelstein coordinates allowed already to extract non-bidiagonal uncharged~\cite{Babichev:2013una,Babichev:2014oua} and charged~\cite{Babichev:2014fka} black hole solutions in massive gravity.
We take the metric $f$ to be fixed and flat,  but  written in an unusual form,
\begin{equation}
\begin{aligned}\label{f}
	ds_f^2 =&  C^2\left[ - dv^2 +2dvdr +2a\sin^2\theta dr d\phi + \rho^2 d\theta^2 
								+\left( r^2+a^2\right) \sin^2\theta d\phi^2 \right].
\end{aligned}
\end{equation}
This form  can be obtained from the canonical Minkowski line element $ds^2_M = -dt^2 + dx^2 +dy^2 +dz^2$ by the coordinate transformation,
$t=v-r$, $x+iy = (r-ia)e^{i\phi}\sin\theta$, $z=r\cos\theta$ (see, e.g. \cite{Visser:2007fj}), and by the subsequent replacement $r\to C r$, $v\to C v$, $a\to C a$.

We will show now that (\ref{g}) and (\ref{f}) are the solution of the massive gravity equation~(\ref{Eg}) with an appropriately chosen conformal factor $C$ and the bare cosmological constant $\Lambda_g$.
First of all, since the metric $g$ coincides with the Kerr solution, we have
$$G_{\mu\nu} = 0.$$
Thus we have to ensure  that the r.h.s. of (\ref{Eg}) is also zero.
For the choice~(\ref{g}) and (\ref{f}) the matrix $\mathcal{K}^\mu_{\phantom{\mu}\nu}$ takes a rather simple form,
with each of the diagonal elements equal to $(1-C)$ and the only non-vanishing non-diagonal ones read, 
$$\mathcal{K}^r_{\phantom{\mu}t} = \frac{C r_g r}{2 \rho^2}, \;
\mathcal{K}^r_{\phantom{\mu}\phi} = \frac{C a r_g r \sin^2\theta}{2 \rho^2}.$$
With the above expression, after lengthy, but straightforward calculation we find from~(\ref{T}),
\begin{equation}\label{Tmn}
T^{\mu}_{\phantom{\mu}\nu} =
\left(
\begin{array}{cccc}
  \lambda_{g} & 0 & 0 & 0 \\
 T^r_{\phantom{r}v} &    \lambda_{g} & 0 &  T^r_{\phantom{r}\phi } \\
 0 & 0 &   \lambda_{g} & 0 \\
 0 & 0 & 0 &     \lambda_{g}
\end{array}
\right),
\end{equation}
where we denoted,
\begin{eqnarray}\label{lambdag}
	  \lambda_{g} =-(C-1)\left(\beta (C-1)^2-3 \alpha  (C-1)+3\right),
\end{eqnarray}
and 
\begin{eqnarray}\label{Toff}
	\begin{aligned}
	T^r_{\phantom{r}v} & = - \left(\beta  (C-1)^2-2 \alpha  (C-1)+1\right)  \frac{C r_g r}{2\rho^2},\\
	T^r_{\phantom{r}\phi} & = - \left(\beta  (C-1)^2-2 \alpha  (C-1)+1\right) \frac{C a r_g r \sin^2\theta}{2 \rho^2},
	\end{aligned}
\end{eqnarray}
and introduced the following notations,
\begin{equation}\label{ab}
\alpha\equiv 1+\alpha_3,\ \beta \equiv \alpha_3+\alpha_4.
\end{equation}
Note that the diagonal part of (\ref{Tmn}) acts as an effective cosmological constant, in dimensionful units it reads $\Lambda^{\rm{eff}}_g = -M_P^2 m^2 \lambda_g$.
This part can be cancelled by an appropriate choice of $\Lambda_g$, namely, 
\begin{equation}\label{Ll}
\Lambda_g = -(C-1)\left(\beta (C-1)^2-3 \alpha  (C-1)+3\right).
\end{equation}
On the other hand, both non-diagonal pieces of (\ref{Tmn}) cancel for $C$, satisfying the relation,
\begin{equation}\label{C}
\beta  (C-1)^2-2 \alpha  (C-1)+1 = 0.
\end{equation}
Thus, we have just shown that the metrics (\ref{g}), (\ref{f}) with $C$ given by (\ref{C}) are the rotating Kerr solution of massive gravity model (\ref{action})
with $\Lambda_g$ given by (\ref{Ll}). 

It is worth to discuss the choice of the ansatz. 
The original form of the Kerr line element~(\ref{g}) for $g$ 
and the Kerr-like line element~(\ref{f}) for the fixed metric is essential for establishing the existence of the solution.
Of course, after we have found the solution in this form, it is not difficult 
 to consider it in
another coordinate system. 
However, in a new coordinate system the solution would  have been hardly possible to guess. 
For example, via the  coordinate change 
\begin{equation}\label{coord1}
	v\to t+r +r_g \int \frac{rdr}{\Delta}, \; \phi \to -\phi -a \int \frac{dr}{\Delta},
\end{equation}
where 
\begin{equation}
\Delta = r^2-r_g r+a^2,
\end{equation}
Eq.~(\ref{f}) takes the familiar Boyer-Lindquist form,
\begin{equation}
\begin{aligned}\label{gBL}
	ds_g^2  = & -\left( 1- \frac{r_g r}{\rho^2}\right)dt^2 - \frac{2 a r_g r \sin^2\theta}{\rho^2} dt d\phi + \frac{\rho^2 dr^2}{\Delta}\\
			&+ \rho^2 d\theta^2 + \left( r^2+a^2 + \frac{ a^2 r_g r \sin^2\theta}{\rho^2} \right)\sin^2\theta d\phi^2.
\end{aligned}
\end{equation}
However, in these coordinates the metric $f$ does not look particularly simple,
\begin{equation}
\begin{aligned}\label{fBL}
	\frac{ds_f^2}{C^2}  = & -dt^2 - \frac{2 r_g r}{\Delta}dt dr + \frac{\Delta -r_g r}{\Delta^2} \rho^2 dr^2\\
		& +\frac{2 a r_g r \sin^2\theta}{\Delta}dr d\phi + \rho^2 d\theta^2 + (a^2+r^2)\sin^2\theta d\phi^2.
\end{aligned}
\end{equation}
The point is that while the transformation~(\ref{coord1}) brings the metric $g$ to the Boyer-Lindquist form, at the same time the metric $f$ becomes very complicated. 
One can do the opposite, 
i.e. perform the change of coordinates ~(\ref{coord1}) with $r_g=0$ to bring the metric $f$ to the simple canonical diagonal form at large $r$,
\begin{equation}\label{fBL0}
	\frac{ds_f^2}{C^2} = -dt^2 +dr^2 + r^2 d\theta^2 + r^2\sin^2\theta d\phi^2,
\end{equation}
where we neglected terms $\mathcal{O}(r^{-2})$. In the same approximation, the metric $g$ takes the following form in this asymptotically unitary gauge,
\begin{equation}
\begin{aligned}\label{gBL0}
	ds_g^2 = & -\left( 1- \frac{r_g}{r}\right)dt^2 + \frac{2 r_g}{r}dtdr -\frac{2 a r_g \sin^2\theta}{r} dt d\phi +\left( 1+ \frac{r_g}{r}\right)dr^2\\
			& -\frac{2 a r_g \sin^2\theta}{r}drd\phi + r^2 d\theta^2 + r^2 \sin^2\theta d\phi^2.
\end{aligned}
\end{equation}
Eq.~(\ref{gBL0}) is the approximate solution (up to $\mathcal{O}(r^{-2})$ terms) in the unitary gauge~(\ref{fBL0}). 
This would be not an obvious guess to start with when trying to find a solution in the unitary gauge. 

Indeed, let us start with the unitary gauge~(\ref{fBL0}) and try guess the solution for the physical metric~$g$. 
An obvious choice would be the Kerr solution in the Boyer-Lindquist coordinates. Thus we take~(\ref{gBL}) and make an expansion in powers ($1/r$) up to $\mathcal{O}(1/r^2)$.
This gives,
\begin{equation}
\label{gBLe}
	ds_g^2 =  -\left( 1- \frac{r_g}{r}\right)dt^2  -\frac{2 a r_g \sin^2\theta}{r} dt d\phi +\left( 1+ \frac{r_g}{r}\right)dr^2
			 + r^2 d\theta^2 + r^2 \sin^2\theta d\phi^2.
\end{equation}
Note the absence of $dt dr$ and $drd\phi$ terms of order $\mathcal{O}(r^{-1})$ in~(\ref{gBLe}) as 
 compared to~(\ref{gBL0}). The metric~(\ref{gBLe}) is obviously an approximate 
solution to the vacuum Einstein equations~$G_{\mu\nu}=0$, since it is an expansion of the Kerr solution at large $r$. 
The calculation of the mass energy-momentum tensor gives $T^{\mu}_{\phantom{\mu}\nu} \propto \delta^\mu_\nu + \mathcal{O}(r^{-2})$, provided that (\ref{C}) holds.
This means that with  an appropriate choice of $\Lambda_g$  (in fact the one made in (\ref{Ll})), the metrics (\ref{fBL0}) and (\ref{gBLe}) are an approximate solution (as $r\to\infty$) 
to massive gravity equations. 
This fact may suggest that apart from the solution~(\ref{f}) and (\ref{g}), or alternatively (\ref{fBL}) and (\ref{gBL}), there is another exact solution, 
approximated by (\ref{fBL0}) and (\ref{gBLe}) at $r\to \infty$.
Unfortunately, we could not find it. 
Our naive guess would be to take $f$-metric to be exactly (\ref{fBL0}) and expand (\ref{gBL}) up to the next order. As a result, however, the 
mass energy-momentum tensor still deviates from the effective Lambda-term at $\mathcal{O}(r^{-2})$ order, provided (\ref{C}) is satisfied. 
It is interesting to mention that the terms of the order $\mathcal{O}(r^{-3})$ are cancelled if the additional condition is satisfied,
\begin{equation}\label{ab}
	\alpha^2 = \beta.
\end{equation}
In fact, in the case of spherically symmetry~\cite{Babichev:2014fka}, this condition allows to make {\it independent} coordinate transformation $v\to v(v,r)$, $r\to r(v,r)$ for each metric 
(as opposite to the {\it common} coordinate transformations a priori leaving the Lagrangian invariant in massive gravity). 
One may expect a similar situation in the case of rotation. 
This, however, does not happen. Indeed, for the metrics (\ref{fBL0}) and (\ref{gBL0}) the expansion of the mass energy-momentum tensor in powers of $1/r$, provided (\ref{C}) and (\ref{ab}) are satisfied,
gives $T^{\mu}_{\phantom{\mu}\nu} \propto \delta^\mu_\nu + \mathcal{O}(r^{-4})$. So the metrics (\ref{fBL0}), (\ref{gBLe}) are approximate solutions up to $\mathcal{O}(r^{-4})$, but not an exact solution. 

A way to generalise the solution we found~(\ref{f}), (\ref{g}) is to try to find the corresponding Kerr-de-Sitter solution. 
The Eddington-Finkelstein-like coordinates would be a starting point if we follow the logic of our paper. 
However, we were not able to find a convenient Eddington-Finkelstein-like form of the Kerr-de-Sitter metric. 
The appropriate transformation bringing the Kerr-de-Sitter solution from the Boyer-Lindquist coordinates to 
the Eddington-Finkelstein-like line element results in a complicated expression, which is extremely difficult to handle technically. 

It is possible to generalise our result to the case of the bi-gravity extension, i.e. to make the second metric dynamical and non-trivial. 
In this case, an extra term, giving dynamics to metric $f$, should be added to the action~(\ref{action}), resulting in,
\begin{equation}\label{actionf}
S =  M^2_P\int d^4 x \sqrt{-g} \left(\frac{R[g]}{2} + m^2 \mathcal{U} [g,f] - m^2 \Lambda_g  \right) + \frac{\kappa M^2_P}{2}\int d^4 x \sqrt{-f} \left(\mathcal{R}[f] -m^2 \Lambda_f \right),
\end{equation}
where $\mathcal{R}[f] $ is the Einstein-Hilbert term for the metric $f$, $\Lambda_f $ is the bare cosmological constant corresponding to $f$ (to be fixed later),
and $\kappa$ is a dimensionless constant, parametrising the difference in Planck masses for $g$ and $f$ metrics.
Besides the metric equations~(\ref{Eg}), metric $f$ also has dynamics now,
\begin{eqnarray}
	\mathcal{G}^{\mu}_{\phantom{\mu}\nu}  =  m^2 \left(
		 \frac{\sqrt{-g}}{\sqrt{-f}}\frac{ \mathcal{T}^{\mu}_{\phantom{\mu}\nu}}{\kappa} - \Lambda_f \delta^\mu_\nu \right) ,\label{Ef}
\end{eqnarray}
where $\mathcal{G}^{\mu}_{\phantom{\mu}\nu}$ is the Einstein tensor for the $f$ metric and $\mathcal{T}^{\mu}_{\phantom{\mu}\nu}$ is 
the mass energy momentum tensor for $f$, given by
\begin{equation}\label{Tf}
\mathcal{T}^\mu_{\phantom{\mu}\nu} = -T^\mu_{\phantom{\mu}\nu} + \mathcal{U}\delta^\mu_\nu.
\end{equation}
We take the metric $g$ to be again the Kerr metric in the Eddington-Finkelstein-like form~(\ref{g}), as in the case with one dynamical metric.
As for $f$, now we do not assume it to be flat, but of the Kerr form, 
\begin{equation}
\begin{aligned}\label{fdyn}
	ds_f^2 = C^2[
	& -\left( 1- \frac{2 r_f r}{\rho^2}\right)\left( dv+a\sin^2\theta d\phi \right)^2 \\
			& + 2\left( dv+a\sin^2\theta d\phi \right)\left( dr+a\sin^2\theta d\phi \right) 
			+ \rho^2 \left( d\theta^2 +\sin^2\theta d\phi^2 \right)].
\end{aligned}
\end{equation}
which is similar to the metric $g$, Eq.~(\ref{g}), besides the different Schwarzschild radii $r_f$.
The result of the calculation is similar to the case of one dynamical metric we studied above. In particular, the mass energy-momentum tensor entering the $g$-metric equations 
is  given by~(\ref{Tmn}), with the only difference that in the expression for the non-diagonal elements $r_g\to r_g-r_f$.
Therefore the equations of motion for the $g$-metric are satisfied if (\ref{Ll}) and (\ref{C}) are satisfied, as in the case of one dynamical metric.
As for the equations for the metric $f$, the l.h.s. of (\ref{Ef}) is automatically zero for the metric (\ref{fdyn}). 
The off-diagonal part of $\mathcal{T}^\mu_{\phantom{\mu}\nu}$ is cancelled if (\ref{C}) is satisfied by virtue of (\ref{Tf}). The diagonal 
part of (\ref{Ef}) is satisfied if the bare cosmological constant $\Lambda_f$ is tuned to be,
\begin{equation}\label{Llf}
\frac{1}{\kappa C^3}\left( C^3 (1-\alpha+\beta) -3C^2 \beta+ 3C (\alpha+\beta)-2\alpha-\beta -1\right) = \Lambda_f.
\end{equation}
Therefore, Eqs~(\ref{g}) and (\ref{fdyn}) are the solution of the bi-gravity theory (\ref{actionf}) provided that (\ref{C}), (\ref{Ll}) and (\ref{Llf})
are satisfied.

Let us now discuss more concretely Eq.~(\ref{C}).   
Note that not all values of $\alpha$ and $\beta$ give physically acceptable solutions for $C=C(\alpha,\beta)=\frac{\alpha+\beta \pm\sqrt{\alpha^2-\beta}}{\beta}$. 
Indeed, for $\alpha^2 < \beta$ the solutions of (\ref{C}) 
are complex valued. Moreover, there is a one-parameter family, given by the condition $\beta+2\alpha+1=0$,
which gives $C=0$. This solution for $C$ is also unphysical, since the metric $f$ is singular.
It is interesting to make a connection of the results of ~\cite{Hassan:2014vja} with our findings. 
In \cite{Hassan:2014vja} the authors found that for a particular subclass of the bi-gravity model~(\ref{actionf}) (i.e. a particular choice of the coefficients in the action),
the only allowed solutions are those with two proportional metrics. This means that for these particular cases non bidiagonal solutions do not exist, i.e. Eq. (\ref{C}) is not verified.
This particular subclass (called the $\beta_i$ model in \cite{Hassan:2014vja}) is translated as follows in our definitions (see, e.g.~\cite{Babichev:2013usa} for identifications between the two different notations). 
The case $\beta_1$ of \cite{Hassan:2014vja} corresponds to $\alpha_3=-1$ and $\alpha_4=1$ (i.e. $\beta=\alpha=0$). This choice leads to the wrong equality, $1=0$, in (\ref{C}).
The cases $\beta_2$ and $\beta_3$ of \cite{Hassan:2014vja} imply $\alpha_3=-3/2$, $\alpha_4=3/2$ ($\beta=0,\alpha=-\frac{1}{2}$)  and $\alpha_3=-2$, $\alpha_4=3$ ($\beta=1,\ \alpha=-1$). Both cases lead to the solution $C=0$, which is unphysical.
Thus 
the $\beta_i$ models studied in \cite{Hassan:2014vja} indeed imply the absence of non-bidiagonal solutions. 
Interestingly, from our discussion we see that
the range of parameters for which our non-bidiagonal solution does not exist is much wider than the parameters corresponding 
to the particular points of the $\beta_i$ models.

Another interesting aspect discussed in~\cite{Hassan:2014vja} is the pathologies connected to the non-bidiagonal branch of cosmological solutions.
In particular, for this branch of solutions it happens that the linear perturbations of the scalar sector disappear, which signals a pathology in the theory. 
It seems, however, that this particular behaviour is connected to the symmetries of the FRW background. 
Indeed, by studying spherically-symmetric  perturbations of non-bidiagonal black hole solutions we found~\cite{Babichev:2014oua} that, provided
the theory is not fine-tuned to have $\beta=\alpha^2$, 
the helicity-0 perturbations are present and do not show this kind of pathology. 
Therefore we expect that the rotating non-bidiagonal solutions will not show this pathology. Work in this direction is in progress.

A further way to generalise our result is to include charge, i.e. to construct an analogue of the Kerr-Newman metric.
And indeed, it is not difficult to do. Let us consider the bi-gravity model~(\ref{actionf}) (to be more general) and also add the standard Maxwell term,
\begin{equation}\label{maxwell}
	-\frac14 \int d^4 x \sqrt{-g} F_{\mu\nu}F^{\mu\nu}.
\end{equation}
The ansatz for $f$ is the same, (\ref{fdyn}), but instead of (\ref{g}) we take the Kerr-Newman line element~\cite{Newman:1965my} in the Eddington-Finkelstein form, which reads
\begin{equation}
\begin{aligned}\label{gkn}
	ds_g^2 = 
	& -\left( 1- \frac{r_g r -r_Q^2}{\rho^2}\right)\left( dv+a\sin^2\theta d\phi \right)^2 \\
			& + 2\left( dv+a\sin^2\theta d\phi \right)\left( dr+a\sin^2\theta d\phi \right) 
			+ \rho^2 \left( d\theta^2 +\sin^2\theta d\phi^2 \right),
\end{aligned}
\end{equation}
where $r_Q$ is the scale corresponding to the electric charge (to be fixed below in terms of the electric charge).
The calculations in fact do not change much in comparison to the Kerr case, the only difference is that the Einstein tensor is not zero now, 
but it has some contribution to the curvature because of the charge. 
This contribution is cancelled by the non-trivial electric field 
\begin{equation}\label{A}
	A_\mu = Q\left\{\frac{r}{\rho^2},\, 0,\, 0,\, \frac{a r \sin^2\theta}{\rho^2} \right\},
\end{equation}
provided that $r_Q$ is expressed in terms of the electric charge $Q$ as follows
\begin{equation}\label{Q}
\sqrt{2}M_P r_Q  = Q,
\end{equation}
Note that (\ref{A}) also satisfies the Maxwell equation $\nabla_\mu F^{\mu\nu} =0$ with $F_{\mu\nu} = \partial_\mu A_\nu - \partial_\nu A_\mu$.
As a result, we are left with the same conditions (\ref{C}), (\ref{Ll}) and (\ref{Llf}), which ensure the validity of the metric equations~(\ref{Eg}) and (\ref{Ef}).
Thus the metrics (\ref{gkn}) with (\ref{Q}), (\ref{fdyn}) and the electromagnetic potential~(\ref{A}) give a solution of the massive bi-gravity theory (\ref{actionf}) with the Maxwell matter term~(\ref{maxwell}),
provided that the conditions  (\ref{C}), (\ref{Ll}) and (\ref{Llf}) are satisfied.
It is not difficult to see that this is also a solution for the dRGT model with one dynamical metric, in this case we only need to set to zero the mass of the $f$-black hole, $r_f=0$, so that 
(\ref{fdyn}) becomes (\ref{f}), and omit the condition~(\ref{Llf}), since the metric $f$ is non-dynamical.

It is worth to mention that setting the rotation parameter to zero, $a=0$, the metrics~(\ref{gkn}) and (\ref{fdyn}) and the potential~(\ref{A}) become
the solution for the static charged  black hole found in \cite{Babichev:2014fka}.

Let us also mention another (simple) solution. 
It is easy to see that the metrics (\ref{g}) and (\ref{fdyn}) with $C=1$ and $r_f=r_g$ also form a solution of massive gravity equations, 
since in this case the massive energy-momentum tensor is 
automatically zero. Another, less trivial solution is  (\ref{g}) and (\ref{fdyn}) with $r_f=r_g$, but $C\neq 1$. In this case the mass energy-momentum is not zero, 
however, the non-diagonal elements vanish, since they are proportional to $(r_g-r_f)$. 
Thus, to satisfy (\ref{Eg}) and (\ref{Ef}), Eqs.~(\ref{Ll}) and (\ref{Llf}) must be satisfied.

\section{Conclusions and discussions}
In this paper we presented  the first rotating black hole solutions in massive (bi)gravity. 
We first showed how to find the solution in the dRGT theory, with one dynamical metric:
The metrics (\ref{g}), (\ref{f}) form a solution of the massive gravity theory~(\ref{action}), provided that  (\ref{C}) and (\ref{Ll}) are satisfied.

Our key idea was to start with the metric line element in the original Kerr form (\ref{g}), 
while the fiducial metric $f$ is flat and written in the rotating coordinates~(\ref{f}), which is simply the Kerr metric with zero black hole mass.
This form of the metrics allowed us to avoid heavy calculations due to a simple form of the matrix $\sqrt{g^{\mu\alpha}f_{\alpha\nu}} $,
resulting in a rather simple expression for the mass energy-momentum tensor~(\ref{Tmn}).
We found then that the conditions~(\ref{C}) on the conformal factor $C$, and~(\ref{Ll}) on the parameters of the action must be verified
 in order to cancel the off- and on-diagonal terms of the r.h.s. of~(\ref{Eg}).

This solution can also be extended to the case of bi-gravity, when the two metrics are dynamical.
Indeed, we showed that the metrics (\ref{g}) and (\ref{fdyn}) form a solution of massive bi-gravity (\ref{actionf}) provided that (\ref{C}), (\ref{Ll}) and (\ref{Llf}) are satisfied.

It is also possible to further generalise the solution to include the black hole charge. The metric $g$ in this case is given by the Kerr-Newman solution 
in the Eddington-Finkelstein coordinates~(\ref{gkn}), the metric $f$ is the same as in the case of the rotating solution, (\ref{fdyn}), and the electromagnetic potential is given by (\ref{A}).
Provided the validity of~(\ref{C}), (\ref{Ll}), (\ref{Llf}) and (\ref{Q}) (which is simply a definition of $r_Q$), this gives a rotating charged black hole solution for the massive gravity theory~(\ref{actionf})
with the additional Maxwell matter term~(\ref{maxwell}). 

Another class of rotating solutions can be readily given. When the metrics $g$ and $f$ are proportional, i.e. for $r_f=r_g$, 
(\ref{g}) and (\ref{fdyn}) comprise a solution of (\ref{actionf}), provided (\ref{Ll}) and (\ref{Llf}) are satisfied. In this case the off-diagonal terms in the 
mass energy-momentum tensor are zero thanks to the condition $r_g=r_f$.
In particular, this class of solutions includes the most simple one when the two metric coincide, i.e. $C=1$, $r_g=r_f$.
The bare cosmological constants are zero in this case, $\Lambda_g=\Lambda_f =0$.

We also discussed a way to generalise the presented solution to asymptotically de-Sitter space-time, i.e. to obtain an analogue of the Kerr-de-Sitter solution. 
We, however, were not able to find such a solution, in particular because we did not find a simple Eddington-Finkelstein-like form of the Kerr-de-Sitter metric.
Another type of generalisation would be to allow the rotating parameters to differ for the two metrics, however, 
in this case the calculation become extremely heavy and we were not able to get the result. This, of course, does not mean that such a solution does not exist, 
one probably needs to find another technique to compute the mass energy-momentum tensor. These two questions would be interesting to investigate in future.

It is worth to mention the asymptotic behaviour of our solution at spatial infinity, $r\to \infty$. In the case of one dynamical metric, 
by appropriate coordinate change, the metric $f$ can be brought to the canonical flat spherically symmetric line element (up to the conformal factor), see (\ref{fBL0}),
while $g$ in this case has (asymptotically) the form~(\ref{gBL0}). 
On the other hand, a naive guess for the metric~$g$, Eq.~(\ref{gBLe}), which is the expansion of the Kerr metric in the Boyer-Lindquist coordinates, in fact,
gives an asymptotic solution up to $\mathcal{O}(r^{-2})$, provided that (\ref{C}) and (\ref{Ll}) are satisfied. 
This is surprising, since the expansion of our exact solution (\ref{fBL0}) in unitary gauge differs from the 
guessed solution (\ref{gBLe}) already at $\mathcal{O}(r^{-1})$ order. 
Moreover, if we take (\ref{fBL0}) and (\ref{gBL}) to be exact and apply the extra condition~(\ref{ab}), the metric equations are satisfied up to $\mathcal{O}(r^{-4})$.
If this is not a coincidence, 
it may mean that there is another solution, whose asymptotic behaviour is given by (\ref{fBL0}) and (\ref{gBLe}). It would be interesting to continue this study in future works.

We note also that because the metric $g$ of our solution has the Kerr form, it is physically relevant, since in this case 
the motion of bodies in the gravitational field of the rotating black hole (assuming the matter couples only to $g$-metric) is the same as in General Relativity.
The massive gravity theory can be distinguishable from General Relativity on the level of perturbations. We leave this research for future work.

We also discussed our results in view of the recent paper~\cite{Hassan:2014vja}. 
Since the two metrics of our rotating solution are of the Einstein form, our solution belongs to the class of solutions discussed in 
\cite{Hassan:2014vja}.
For the so-called $\beta_i$ models (particular cases of the model (\ref{actionf})) the authors of \cite{Hassan:2014vja} found that 
the only solutions that exist in the class of solutions they considered are those with proportional metrics. 
We confirmed their findings by showing that equation (\ref{C}), that must be satisfied in order for the non-bidiagonal solutions to exist, either cannot be satisfied or gives $C=0$, an unphysical solution.
On the other hand we found that the range of parameters for which (\ref{C}) does not lead to a physical solution is much wider than just the $\beta_i$ models.

Finally, a stability analysis of the found solutions should be addressed along the lines of~\cite{Babichev:2013una,Brito:2013wya,Babichev:2014oua}.
As the analysis of \cite{Babichev:2014oua} has shown, the spherical perturbations of non-bidiagonal spherically-symmetric black hole solutions do not  
show the strong coupling (unless the parameters of the model are fine-tuned) --- a pathology which seems to be a general feature for non-bidiagonal cosmological solutions.
Therefore we do not expect that the strong coupling appears in the spherical modes of these rotating black holes. This work is currently in progress.

{\it Acknowledgments.} 
A.F. thanks LPT for hospitality during various visits.
The work of E.B. was supported in part by the Grant No. RFBR 13-02-00257-a.


\begin{thebibliography}{99}

\bibitem{deRham:2014zqa}
  C.~de Rham,
  arXiv:1401.4173 [hep-th].

\bibitem{Boulware:1973my}
  D.~G.~Boulware, S.~Deser,
  Phys.\ Rev.\ D {\bf 6} (1972) 3368.


\bibitem{deRham:2010ik}
  C.~de Rham, G.~Gabadadze,
  Phys.\ Rev.\ D {\bf 82} (2010) 044020
  [arXiv:1007.0443 [hep-th]];
  C.~de Rham, G.~Gabadadze, A.~J.~Tolley,
  Phys.\ Rev.\ Lett.\  {\bf 106} (2011) 231101
  [arXiv:1011.1232 [hep-th]].
  S.~F.~Hassan, R.~A.~Rosen,
  Phys.\ Rev.\ Lett.\  {\bf 108} (2012) 041101
  [arXiv:1106.3344 [hep-th]];
  S.~F.~Hassan, R.~A.~Rosen,
  JHEP {\bf 1202} (2012) 126
  [arXiv:1109.3515 [hep-th]].

\bibitem{Kluson:2011rt}
  J.~Kluson,
  JHEP {\bf 1206} (2012) 170
  [arXiv:1112.5267 [hep-th]];
  J.~Kluson,
  Phys.\ Rev.\ D {\bf 86} (2012) 044024
  [arXiv:1204.2957 [hep-th]];
  C.~Deffayet, J.~Mourad and G.~Zahariade,
  JCAP {\bf 1301} (2013) 032
  [arXiv:1207.6338 [hep-th]];
  C.~Deffayet, J.~Mourad and G.~Zahariade,
  JHEP {\bf 1303} (2013) 086
  [arXiv:1208.4493 [gr-qc]];
  A.~Golovnev,
  Phys.\ Lett.\ B {\bf 707} (2012) 404
  [arXiv:1112.2134 [gr-qc]].
  

\bibitem{Vainshtein:1972sx}
A.~I.~Vainshtein,
Phys.\ Lett.\ B {\bf 39} (1972) 393.

\bibitem{Babichev:2009us}
  E.~Babichev, C.~Deffayet, R.~Ziour,
  JHEP {\bf 0905} (2009) 098
  [arXiv:0901.0393 [hep-th]];
  E.~Babichev, C.~Deffayet, R.~Ziour,
  Phys.\ Rev.\ Lett.\  {\bf 103} (2009) 201102
  [arXiv:0907.4103 [gr-qc]];
  E.~Babichev, C.~Deffayet, R.~Ziour,
  Phys.\ Rev.\ D {\bf 82} (2010) 104008
  [arXiv:1007.4506 [gr-qc]];
  E.~Babichev, C.~Deffayet and R.~Ziour,
  Int.\ J.\ Mod.\ Phys.\ D {\bf 18} (2009) 2147
  [arXiv:0905.2943 [hep-th]];
  L.~Alberte, A.~H.~Chamseddine and V.~Mukhanov,
  JHEP {\bf 1012} (2010) 023
  [arXiv:1008.5132 [hep-th]];
  G.~Chkareuli and D.~Pirtskhalava,
  Phys.\ Lett.\ B {\bf 713} (2012) 99
  [arXiv:1105.1783 [hep-th]];
  E.~Babichev and M.~Crisostomi,
  Phys.\ Rev.\ D {\bf 88} (2013) 084002
  [arXiv:1307.3640];
  S.~Renaux-Petel,
  JCAP {\bf 1403} (2014) 043
  [arXiv:1401.0497 [hep-th]].
  L.~Berezhiani, G.~Chkareuli and G.~Gabadadze,
  Phys.\ Rev.\ D {\bf 88} (2013) 124020
  [arXiv:1302.0549 [hep-th]];
  L.~Berezhiani, G.~Chkareuli, C.~de Rham, G.~Gabadadze and A.~J.~Tolley,
  Class.\ Quant.\ Grav.\  {\bf 30} (2013) 184003
  [arXiv:1305.0271 [hep-th]].
  
 
 \bibitem{Koyama:2011xz}
  K.~Koyama, G.~Niz and G.~Tasinato,
  Phys.\ Rev.\ Lett.\  {\bf 107} (2011) 131101  [arXiv:1103.4708 [hep-th]];
  K.~Koyama, G.~Niz and G.~Tasinato,
  Phys.\ Rev.\ D {\bf 84} (2011) 064033
  [arXiv:1104.2143 [hep-th]];

 
\bibitem{Babichev:2013usa}
  E.~Babichev and C.~Deffayet,
  Class.\ Quant.\ Grav.\  {\bf 30} (2013) 184001
  [arXiv:1304.7240 [gr-qc]].

\bibitem{Salam:1976as}
  A.~Salam and J.~A.~Strathdee,
  Phys.\ Rev.\ D {\bf 16} (1977) 2668;
  C.~J.~Isham and D.~Storey,
  Phys.\ Rev.\ D {\bf 18} (1978) 1047.
  
\bibitem{Comelli:2011wq}
  D.~Comelli, M.~Crisostomi, F.~Nesti and L.~Pilo,
  Phys.\ Rev.\ D {\bf 85} (2012) 024044
  [arXiv:1110.4967 [hep-th]].

\bibitem{Volkov:2012wp}
  M.~S.~Volkov,
  Phys.\ Rev.\ D {\bf 85} (2012) 124043
  [arXiv:1202.6682 [hep-th]];
  
\bibitem{Berezhiani:2008nr}
  Z.~Berezhiani, D.~Comelli, F.~Nesti and L.~Pilo,
  JHEP {\bf 0807} (2008) 130
  [arXiv:0803.1687 [hep-th]].
  
\bibitem{Nieuwenhuizen:2011sq}
  T.~M.~Nieuwenhuizen,
  Phys.\ Rev.\ D {\bf 84} (2011) 024038
  [arXiv:1103.5912 [gr-qc]].

\bibitem{Berezhiani:2011mt}
  L.~Berezhiani, G.~Chkareuli, C.~de Rham, G.~Gabadadze and A.~J.~Tolley,
  Phys.\ Rev.\ D {\bf 85} (2012) 044024
  [arXiv:1111.3613 [hep-th]].
  
\bibitem{Babichev:2014fka}
  E.~Babichev and A.~Fabbri,
  JHEP {\bf 1407} (2014) 016
  [arXiv:1405.0581 [gr-qc]].
    
\bibitem{Brito:2013xaa}
  R.~Brito, V.~Cardoso and P.~Pani,
  Phys.\ Rev.\ D {\bf 88} (2013) 064006
  [arXiv:1309.0818 [gr-qc]].
  
 \bibitem{VolkovTasinato} 
  M.~S.~Volkov,
  Class.\ Quant.\ Grav.\  {\bf 30} (2013) 184009
  [arXiv:1304.0238 [hep-th]].
  G.~Tasinato, K.~Koyama and G.~Niz,
  Class.\ Quant.\ Grav.\  {\bf 30} (2013) 184002
  [arXiv:1304.0601 [hep-th]].

\bibitem{Babichev:2013una}
  E.~Babichev and A.~Fabbri,
  Class.\ Quant.\ Grav.\  {\bf 30} (2013) 152001
  [arXiv:1304.5992 [gr-qc]].
  
\bibitem{Brito:2013wya}
  R.~Brito, V.~Cardoso and P.~Pani,
  Phys.\ Rev.\ D {\bf 88} (2013) 023514
  [arXiv:1304.6725 [gr-qc]].
  
\bibitem{Babichev:2014oua}
  E.~Babichev and A.~Fabbri,
  Phys.\ Rev.\ D {\bf 89} (2014) 081502
  [arXiv:1401.6871 [gr-qc]].

\bibitem{Kodama:2013rea}
  H.~Kodama and I.~Arraut,
  Prog. Theor. Exp. Phys. (2014) 023E02
  [arXiv:1312.0370 [hep-th]].
     
\bibitem{Kerr:1963ud}
  R.~P.~Kerr,
  Phys.\ Rev.\ Lett.\  {\bf 11} (1963) 237.
  
\bibitem{Visser:2007fj}
  M.~Visser,
  arXiv:0706.0622 [gr-qc].
  
\bibitem{Newman:1965my}
  E T.~Newman, R.~Couch, K.~Chinnapared, A.~Exton, A.~Prakash and R.~Torrence,
  J.\ Math.\ Phys.\  {\bf 6} (1965) 918.
 
\bibitem{Hassan:2014vja}
  S.~F.~Hassan, A.~Schmidt-May and M.~von Strauss,
  arXiv:1407.2772 [hep-th].
 
 \end{thebibliography}
\end{document}